\documentclass[reprint,aps,prb,amsmath,amssymb]{revtex4}

\usepackage[T1]{fontenc}
\usepackage[]{graphicx}
\usepackage{subfigure}
\usepackage{times}
\usepackage{bm}
\usepackage{color}

\newcommand{\comment}[1]{}

\newcommand{\be}{\begin{equation}}
\newcommand{\ee}{\end{equation}}
\newcommand{\bea}{\begin{eqnarray}}
\newcommand{\eea}{\end{eqnarray}}
\def\mean#1{\left< #1 \right>}

\begin{document}

\title{Ultrafast exciton relaxation in monolayer transition metal dichalcogenides}

\author{A. Thilagam}
\email[]{thilaphys@gmail.com}

\affiliation{Information Technology, Engineering and Environment,\\ 
University of South Australia, Australia
 5095.}
%\today
%%%%%%%%%%%%%%%%%%%%%%%%%%%%%%%%%%%%%%%%%%%%%%%%%%%%%%%%%%%%% 
\begin{abstract}
We examine a mechanism by which excitons undergo ultrafast 
relaxation  in  common monolayer 
  transition metal dichalcogenides. It is shown that
at   densities $\approx$ 1$\times$ 10$^{11}$ cm$^{-2}$
and  temperatures $ \le 60$ K, excitons in well known
monolayers  (MoS$_2$,  MoSe$_2$, WS$_2$ and WSe$_2$)
exist as point-like structureless electron-hole quasi-particles.
We evaluate the  average rate of exciton energy relaxation due to
acoustic phonons via the deformation potential and the piezoelectric
coupling mechanisms and examine the effect of spreading of the excitonic wavefunction
into the region perpendicular to the 
monolayer plane. Our results show that the exciton relaxation   rate 
is enhanced with increase in the  exciton temperature, while it is decreased
with increase in the lattice temperature. 
Good agreements with  available experimental data
are obtained when  the calculations are extrapolated to room temperatures. 
A unified approach taking into account the deformation potential
and piezoelectric coupling mechanisms shows 
that exciton relaxation induced by  phonons is as significant as
defect assisted scattering and  trapping of excitons by surface states
 in monolayer   transition metal dichalcogenides.
%\keywords{Excitons; Polarons; Exciton-phonon interaction;
\end{abstract}

\maketitle

%Pacs: 81.05.Hd, 72.10.-d, 72.20.-i, 72.80.Jc
% 71.35 +z, 71.38 +i, 73.40 -c

\section{Introduction}
In recent years, several works have examined the occurrences of large  exciton binding energies
 \cite{ugeda2014giant,hill2015observation,makatom,chei12,komsa2012effects,thiljap} as well as
exciton mediated ultrafast processes  \cite{mannebach2014ultrafast,konabe2014effect,wang2014ultrafast}
that are enhanced in atomically thin layered transition metal dichalcogenides
 MX$_2$ (M = Mo, W, Nb, and X = S, Se) \cite{wilson,wang,he2014stacking}.
The monolayer Molybdenum and Tungsten dichalcogenides (MoS$_2$,  MoSe$_2$, WS$_2$ and WSe$_2$)
generally have  similar lattice parameters  and electronic configurations \cite{liu2015electronic,mann20142}.
Transition metal dichalcogenides   are known to undergo a
crossover from indirect band gap (bilayer) to direct gap in the
monolayer configuration \cite{makatom} where there is strong   photoluminescence.
The intrinsic band gap range from 0.5 eV to 2 eV, depending on the material composition and number of 
layers in the material system. Specifically, the  confinement of charge carriers within a monolayer space results in
an enhanced  interaction of the exciton with  light, a desirable property that
can be exploited to fabricate highly sensitive  photodetectors and sensor devices. 

The study of exciton based   quantum dynamical processes is important
as excitonic interactions underlie  the unique
optoelectronic properties \cite{kumar2014tunable,komsa2012effects,qiu2013,jariwala2014emerging,qiu2015nonanalyticity}
of monolayer dichalcogenides.
Several  applications  \cite{lembke2015single,pospischil2014solar,
zhang2014m,tsai2014monolayer,radisavljevic2011integrated,yoon2011good}
can be linked to the  rich  many-body effects of excitons
in low-dimensional transition metal dichalcogenides.
The  long lifetimes of  excitons in transition metal dichalcogenides
enable interaction with additional degrees of freedom  linked to the unique   optical selection rules of the 
momentum valleys \cite{xiao2012coupled,suzuki2014valley} of
two-dimensional  MX$_2$ materials. This  enables tuning of the   coupling strength of  the valley to spin degrees of freedom.
which results in  desirable  valleytronics properties \cite{mai2013many} essential for  high speed logic devices
A comprehensive 
understanding of the quantum dynamical processes of two-dimensional
excitons  is needed to fully exploit the  unique properties 
of monolayer transition metal dichalcogenides for  electrooptical and valleytronic devices

Exciton dynamics in monolayer and few-layer MoS$_2$ 2D
crystals  examined using femtosecond transient absorption spectroscopy and
microscopy techniques \cite{shi2013exciton} show
a  highly enhanced intraband relaxation rate $\le$ 0.5 ps in the
 monolayer configuration compared to 20 ps in the bulk material.
This enhanced relaxation rate was  attributed to increased defect
assisted scattering and  trapping of excitons by surface states \cite{shi2013exciton}.
The rapid capture rates of excitons  
 by mid-gap defects from Auger processes was estimated to be about
less than a picosecond to a few picoseconds in monolayer metal dichalcogenides
 \cite{wang2015fast}.  The capture rates which display both 
linear and  quadratic dependences on the exciton density
arises primarily from the large
overlap of the defect state wavefunction  with the conduction
or valence band Bloch state wavefunctions.
The exciton capture rates are almost  comparable to the 
exciton-exciton annihilation (EEA) rates \cite{sun2014observation,konabe2014effect} that
becomes  dominant as a result of the 
 reduced Coulomb screening of spatially confined charge carriers
 in  layered transition  metal dichalcogenides.

The layered structures of metal chalcogenides give rise to a variety of phonon modes
that interact via  short range and long range forces with
unique thickness dependences of  phonons in layered transition  metal dichalcogenides
  \cite{lee,zhao13inter,chakraborty2013layer}.
There exist two  Raman active modes, $A_{1g}$ and $E_{2g}^1$ 
 which exhibit frequency shifts  with layer thickness
in  MoS$_2$  \cite{lee,mol11phonons,zhao13inter}. The $A_{1g}$ mode is associated with 
the out-of-plane displacement of S atoms   while the $E_{2g}^1$ 
is associated with the in-plane but opposing displacements of Mo and S atoms \cite{lee,mol11phonons}.
Photoluminescence studies of single-layer MoS$_2$ \cite{korn2011low} reveal
a long-lived component  due to 
exciton-phonon scattering, which highlights the importance of exciton interaction
with phonons. The occurrence  of  interface and  confined
slab phonon modes \cite{hai1993electron} and
 dimensionality dependent exciton-phonon interactions 
in  low-dimensional material systems \cite{thilagam1997exciton} has not been 
fully examined in monolayer metal dichalcogenides.
To this end, the possibility of an enhanced exciton relaxation  arising from
phonon assisted mechanism  remains  to be explored in 
few-layer transition metal dichalcogenide systems.
In this study we consider the critical role played by  phonons  during
exciton relaxation processes  and   compare computed rates of
exciton energy relaxation  with estimates of enhanced exciton 
relaxation rates associated with surface defects  \cite{shi2013exciton,wang2015fast}
in monolayer and few-layer samples.

In the  monolayer  transition-metal dichalcogenides, 
a direct band gap between the  conduction and
valence band extrema occurs at the $K$ point. While an energy minimum 
occurs  at the $Q$ point along the $K-\Gamma$ axis at the lowest conduction
band, energy is maximized at the  $\Gamma$ point of the valence band   \cite{jin2014intrinsic}.
Charge carriers are generally located in the $K, K'$  conduction band valleys which are 
approximately parabolic for energies less than 300 eV \cite{chei12}.
The  direct excitonic   transitions are enhanced  at the  two
extremal locations of the $K$ point in the monolayer configuration.
Newly formed quasi-two-dimensional excitons that possess a  finite center-of-mass motion
momentum wave vector   undergo one of  four main  processes : (1) jump to
 higher energy band states via absorption
of phonons, (2) relax to lower kinetic energies via
emission of phonons, (3) decay non-radiatively  or radiatively to the continuum
state and lastly (4)  decay into free electron and holes that 
subsequently relax via phonon
emission to lower energy levels. 

In this study,
we consider that newly formed electron-hole
pairs lose energy via emission of  longitudinal-acoustic  (LA) and/or transverse-acoustic (TA) phonons
via the  deformation potential coupling channel.
The exciton relaxation is initiated by
deformations of the lattice potentials by acoustic phonons
which is applicable at the low temperature range.
The low dimensional excitons are considered to  undergo
further energy relaxation along the exciton dispersion curves before
reaching a minimum in exciton energy. 
The temperature range of 5 $-$ 70 K considered here therefore excludes the possibility of
exciton relaxation via  optical phonon modes as these transitions involve higher energies.
Piezoelectricity  which arises from the  linear coupling
between the electrical polarization and
mechanical strain within a crystal  have  reasonable 
strengths in monolayer metal dichalcogenides \cite{alyoruk2015,yin2014observation,
duerloo2013,reed2015}.
The    two-dimensional hexagonal crystal structure
lacks an inversion symmetry, hence there occurs a strain field  which gives rise to 
the piezoelectric properties in transition metal dichalcogenides.
We therefore include the  scattering of exciton due to the piezoelectric coupling  via  acoustic phonons   in this study. 

The paper is organized as follows. In
Sec. \ref{exwave}, we present the  theoretical 
form of exciton wavefunction which is suitable for modeling the
correlated electron-hole pair in
 monolayer transition metal dichalcogenides.
In Sec. \ref{distr}, we discuss the Maxwell-Boltzmann and
 Bose-Einstein  exciton distributions, and 
 justify our choice of a suitable
exciton  distribution  based on the material
properties of the monolayer dichalcogenides given in Table-I.
We also discuss the conditions required  for a statistically degenerate system  of excitons
to  occur in monolayer systems. In Sec. \ref{opera} we present an explicit form of a 
quasi-two dimensional exciton-acoustic phonon
interaction  operator applicable in monolayer systems.
This operator is used to derive  the average rate of exciton energy relaxation,
$\mean{\frac{dE}{dt}}$ in Sec. \ref{relaxpi}.
Numerical results are presented for the monolayer   transition metal dichalcogenides, MoS$_2$,  MoSe$_2$, WS$_2$ and WSe$_2$
and differences in their relaxation properties are  analyzed  in Sec. \ref{relaxpi}.
 The theory of exciton relaxation due to the  piezoelectric coupling mechanism is 
presented in Sec. \ref{pie} along with analysis of numerical results based
on the piezoelectric properties of transition metal dichalcogenides.
Lastly, conclusions are summarized in Sec. \ref{con}.

\section{\label{exwave} Exciton wavefunction in the monolayer transition metal dichalcogenides}

In transition metal dichalcogenides (MX$_2$) 
the intra-layers of the M metal  planes are held between the chalcogen  X  atomic planes
by covalent bonding. The inter-layers  can be easily separated into distinct layers \cite{cho2008,mouri2013}
of just a few  atomic layers.
A hexagonally ordered plane of metal atoms sandwiched between two other hexagon planes of chalcogen
atom represents a single monolayer which we examine using
a  quasi two-dimensional space. This is justified
as the motion of the exciton is mostly confined 
 within the parallel two-dimensional $XY$  layers of the atomic planes and there is restricted exciton motion 
in the $z$ direction  perpendicular to the monolayer plane.
 Accordingly, the  exciton is represented by a  quasi-two dimensional wavefunction \cite{thilagam2015excitonic},
and for simplicity we introduce the  two-band  approximation involving
the  lowest  electron subband and highest hole subband structure.
The   exciton  state  vector denoted by $|\alpha,{\bf K} \rangle$ in the presence
of phonons can  then  be written as
\be
|\alpha = 1s,{\bf K} \rangle =
{v_o \over L} \sum_{\bf r_e,r_h} \exp (i{\bf K.R}) \;
\Psi_{1s}({\bf r_e-r_h},z_e,z_h) \;
a_{1,{\bf r_e}}^{\dagger} \; a_{0,{\bf r_h}}^{}  
|0,n \rangle ,
\label{exk}
\ee
where $v_o$ is the volume of the unit cell
and $L$ is a quantized length of the lattice space.  
We restrict this  study to the 
the lowest $1s$ exciton state without great loss in generality.  
The co-ordinate of
the centre of mass of an exciton, ${\bf R}$ is given by
\bea
\label{cmR}
{\bf R} & = & \alpha_e \;{\bf r_e}+ \alpha_h \;{\bf r_h}, \\
\alpha_e & = & \frac{m_e}{m_e+m_h}, \quad \alpha_h = \frac{m_h}{m_e+m_h} ,
\label{mR}
\eea
where $m_e$  ($m_h$ ) is the effective mass of the electron (hole).
The position vectors and wave  vectors are decomposed into components parallel and perpendicular
to the monolayer plane as ${\bf r} = (r_\|,z)$ and ${\bf k} = (k_\|,k_z)$ respectively.
The   creation operator of an  electron  in  the conduction band at  position $r_e$ 
is denoted by  $a_{1,{\bf r_e}}^{\dagger}$.
The annihilation operator of an electron in the valence band
at position $r_h$   is denoted by $a_{0,{\bf r_h}}^{}$.
The state  $|0,n \rangle$ in Eq.\ref{exk} is given by
\begin{equation}
|0,n\rangle=|0\rangle\;|n\rangle
\label{ph}
\end{equation}
where  $|0\rangle$  is   the electronic vacuum
state  of  the  system  that represents
completely filled valence bands  and empty conduction bands.
The occupation number of phonons
with wavevector $\bf q$ is given by
\begin{equation}
|n \rangle = |n_{\bf 1}\;,\;n_{\bf 2},...n_{\bf q} \rangle
\end{equation}

In Eq.\ref{exk},  $\Psi_{1s}({\bf r_e-r_h},z_e,z_h)$ denotes
the  $1s$th state exciton envelope  function   in the
 in the monolayer configuration.
This state has the  variational envelope function which appear 
as  \cite{taka2}
\begin{equation}
\Psi_{\rm 1s}(\rho, z_e, z_h) = \mathcal{N} \exp \left [ -\left(\gamma^2 \rho^2+ \beta^2 (z_e-z_h)^2 \right)^{1/2} \right ]
 \cos[\frac{\pi z_e}{L_w}]  \cos[\frac{\pi z_h}{L_w}],
\label{exr}
\end{equation}
where $\mathcal{N}$ is the normalization constant and 
$\rho= |\bf r_e-r_h|$ is the relative separation of the electron-hole
pair in the monolayer plane. We have assumed that the excitonic wavefunction undergoes
a discontinuous  transition to zero beyond the region $|z_e|  < \frac{L_w}{2}$
and $|z_h| < \frac{L_w}{2}$, where 
$L_w$ is the average displacement of  electrons and holes in  the $z$ direction.
perpendicular to the monolayer surface.
The  parameters $\beta$ and $\gamma$  are determined
by  minimalizing the energy of the exciton in the presence of phonons, a task that is  not
numerically trivial. We therefore treat $\beta$ and $\gamma$
as adjustable parameters  that appear as  inverses of the
confinement lengths parallel and perpendicular to the monolayer plane.
For the exact two dimensional case, $\beta L_w$ = 0.

We  express the  operators  of  Eq.\ (\ref{exk}) in  the Bloch representation
and convert the  summation over $\bf r_e$  and $\bf r_h$ into integral form and  obtain
\begin{equation}
|\alpha=1s,{\bf K} \rangle = \sum_{\bf k,k'}
\Phi_{1s}({\bf k,k',K})\; \delta_{\bf k-k',K} \; a_{1,{\bf k}}^{\dagger}
\;a_{0,{\bf k'}}^{} |0,n \rangle
\label{istate}
\end{equation}
where $\bf k$ and $\bf k'$ are the wavevectors
of electron and hole respectively and
\begin{equation}
\Phi_{1s}({\bf k,k',K}) =
{1 \over L^2}\int
d^{2} r\int dz_{e}\int dz_{h} \; \Psi_{1s}({\bf r_e-r_h},z_e,z_h)
\exp[i(\alpha_e{\bf K-k}).{\bf r}-i{\bf k_z}z_e+i{\bf k'_z}z_h],
\label{phi}
\end{equation}
where $\alpha_e$ and  $\alpha_h$ are specified in Eq.\ref{mR}.

\section{Maxwell-Boltzmann versus Bose-Einstein  exciton distributions \label{distr}}

The results of exciton relaxation kinetics is very much dependent on the type of
exciton distribution employed during  the modeling process. A widely used distribution
 is the Maxwell-Boltzmann function which appear as
\be
f_{mw}(E) = \exp (- k_b T_{ex}),
\label{boltz}
\ee
where  $T_{ex}$  denotes the exciton temperature and $k_b$ is the boltzmann constant.
The total number of excitons  can be easily evaluated as
\be
N =  \sum_{\bf K_\|} \; f_b(E({\bf K_\|})) =  \frac{ M k_b T_{ex}L^2}{2 \pi \hbar^2},
\label{tnum}
\ee
where $L$ is the quantization length and $M$ is the exciton mass.
 The distribution in Eq.\ref{tnum} is
applicable to the  classical model  of quasi-equilibrium of low dimensional excitons.
The  Maxwell-Boltzmann statistical distribution  of excitons is suited for 
the temperature range, $T \gg T_0$ where the degeneracy temperature $T_0$  is given by \cite{ivanov1999bose}
\be
\label{crosst}
T_0 = \left( \frac{2 \pi \hbar^2 n_{ex}}{g M k_b } \right),
\ee
where $n_{ex}$ is the exciton density, and the  degeneracy factor $g$ = 4, 
expressed a product of the spin and valley degeneracy factors \cite{kaasbjerg12}.

The exciton distribution function based on the 
Bose-Einstein distribution is given by
\bea
\label{bec1}
f_b(E) &=&\left[ \exp \left( \frac{E - \mu_{ex}}{K_B T_{ex}}\right ) - 1 \right]^{-1},\\
\label{bec2}
\mu_{ex} &=& K_B T_{ex} \ln \left[1-\exp \left(- \frac{2 \pi \hbar^2 n_{ex}}{g M k_b T_{ex}} \right) \right],
\eea
where the quasi-two dimensional exciton chemical potential $\mu_{ex}$ is dependent on 
the exciton temperature $T_{ex}$ and exciton density  $n_{ex}$ \cite{ivanov1999bose}. 
 We evaluate the total number of excitons  as
\bea
\nonumber
N &=& \sum_{\bf K_\|} \; f(E({\bf K_\|}))\\
 &=& \frac{ M k_b T_{ex}L^2}{\pi \hbar^2} \; \log [1-\exp(-\frac{\mu_{ex}}{k_b T})]
\label{tnume}
\eea
From Eq. \ref{tnume}, it is seen that the 
 Bose-Einstein  distribution of excitons  is the most appropriate model
at low  temperatures for which
there exist a   large  exciton population  with  small wavevectors.

\begin{figure}[htp]
  \begin{center}
    \subfigure{\label{figa}\includegraphics[width=6cm]{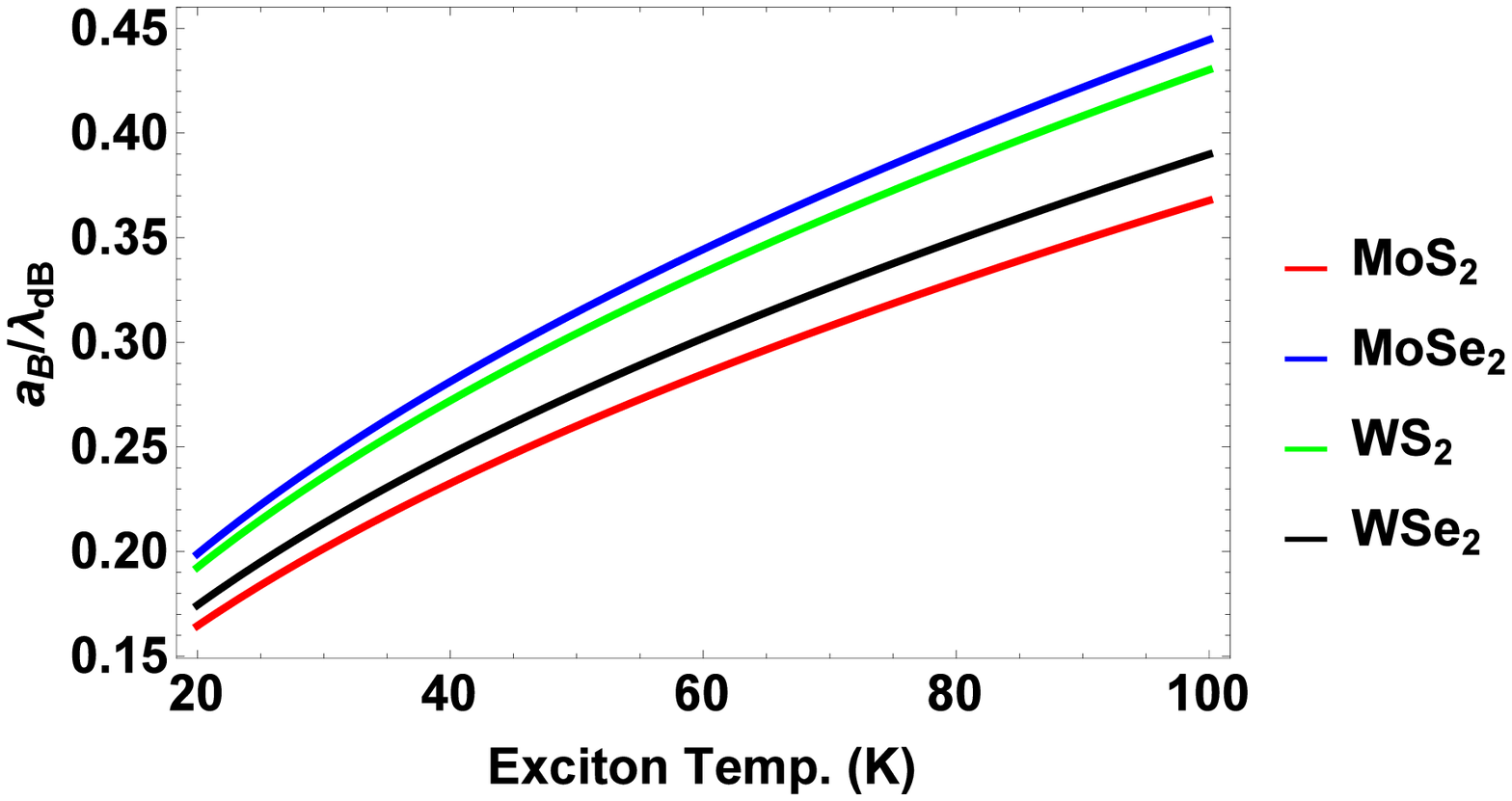}}\vspace{-1.1mm} \hspace{1.1mm} 
\subfigure{\label{figb}\includegraphics[width=4.85cm]{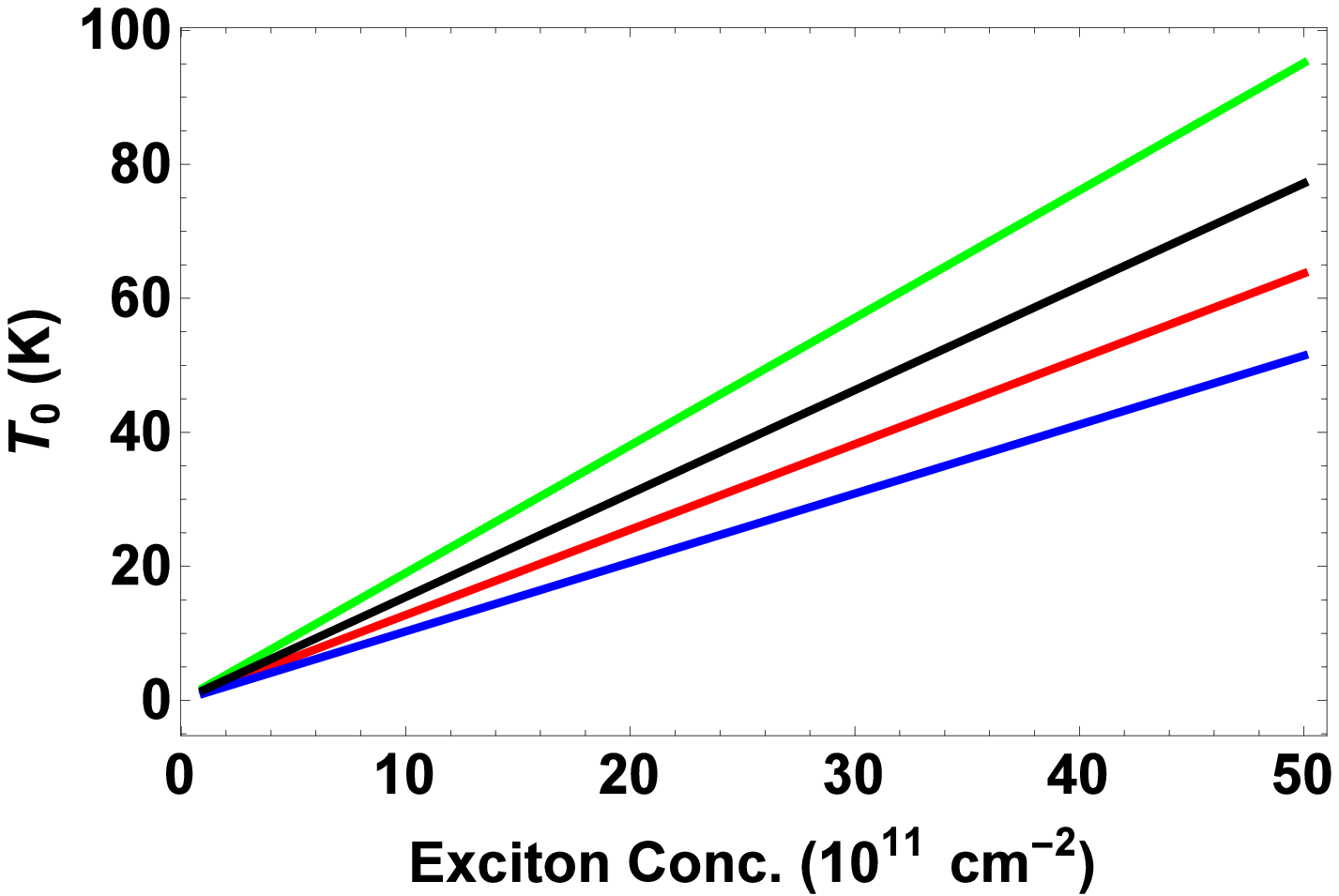}}\vspace{-1.1mm} \hspace{1.1mm} 
\subfigure{\label{figb}\includegraphics[width=4.85cm]{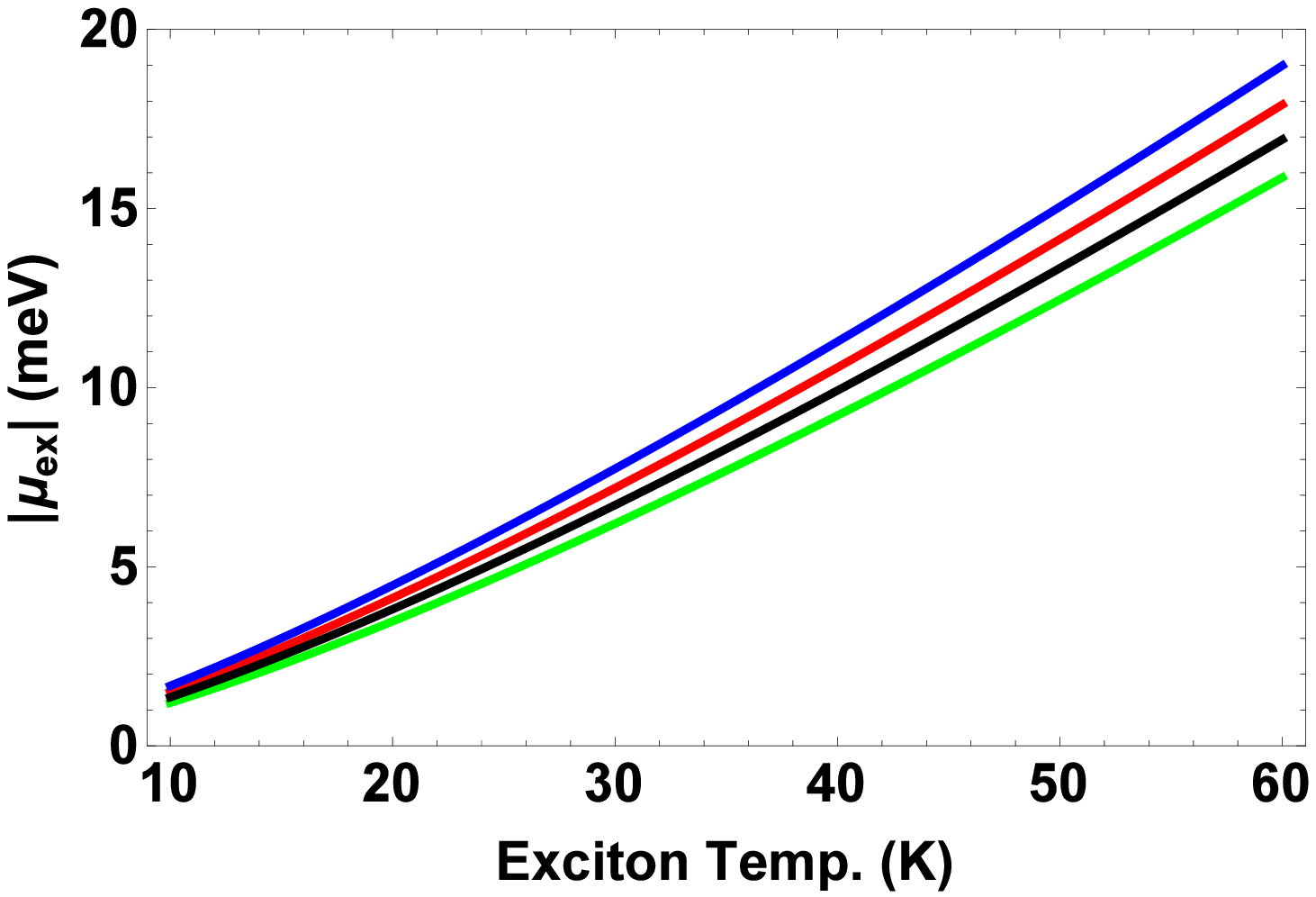}}\vspace{-1.1mm} \hspace{1.1mm} 
 \end{center}
     \caption{(a) The ratio  of exciton Bohr radius to the de Broglie wavelength ( $\frac{a_B}{\lambda_{db}}$)
  as a function of the exciton temperature $T_{ex}$ for the four monolayer   transition metal dichalcogenides, MoS$_2$,  MoSe$_2$, WS$_2$ and WSe$_2$ (left), (b) The degeneracy temperature $T_0$ (Eq.\ref{crosst}) 
 as a function of the exciton  density $n_{ex}$ (middle), (c) The absolute value of the exciton
chemical potential, $|\mu_{ex}|$  as a function of the exciton temperature,
with the exciton  density  fixed at $n_{ex}$ = 1.5 $\times$ 10$^{11}$ cm$^{-2}$ for all the dichalcogenides.
(right).
 }
 \label{pic}
\end{figure}

In  the case of MoS$_2$, we substitute  
the exciton  density,  
$n_{ex}$ = 1$\times$ 10$^{11}$ cm$^{-2}$, and $m_e$ = 0.51, $m_h$=0.58 \cite{jin2014intrinsic}
in  Eq.\ref{crosst} which gives  $T_0$ = 1.3 K. The exciton density of  1$\times$ 10$^{11}$ cm$^{-2}$
corresponds to an inter-particle distance of 316 \AA, which is about 35 times the size
of the exciton radius of 9 \AA \cite{thiljap}. 
Thus  at typical exciton 
temperatures greater than  10 K, the excitons in the  MoS$_2$ monolayer can be considered as well-defined
correlated electron-hole quasi-particles.
Excitons can also be modeled as point-like structureless systems  provided
 the  exciton de Broglie wavelength ($\lambda_{db} = \hbar/\sqrt{2 m T}$)
far exceeds the exciton Bohr radius $a_B$. Based on the material properties
of the monolayer   transition metal dichalcogenides, MoS$_2$,  MoSe$_2$, WS$_2$ and WSe$_2$
(given in Table-I), the ratio  $\frac{a_B}{\lambda_{db}}$ is 
plotted as a function of the exciton temperature $T_{ex}$
in  Fig. \ref{pic}a. There are subtle differences due to varying material properties
of the four types of monolayer systems. In general,
the Maxwellian  distribution 
appears appropriate  for comparatively low exciton temperatures, 
$T_{ex} \le 60$ K and moderate exciton  densities,  
$n_{ex} \approx 1\times$ 10$^{11}$ cm$^{-2}$ (see Fig. \ref{pic}b). We therefore restrict $T_{ex}$ to the  classical temperature range
and exciton concentration to examine the  exciton relaxation  pathways in this study.

In Fig. \ref{pic}b, the degeneracy temperature $T_0$ (Eq.\ref{crosst}) 
is plotted as a function of the exciton  density $n_{ex}$,
for  MoS$_2$,  MoSe$_2$, WS$_2$ and WSe$_2$.
The   $m_e$ and $m_h$ values used in the calculations are
retrieved from Ref. \cite{jin2014intrinsic} (see Table-I). 
Tungsten sulphide yields the
highest estimate of $T_0$  which can be partly attributed to its comparatively small
exciton mass ($m_e$+ $m_h$). The sulfides possess a   higher $T_0$  than the  selenides.
The results in Fig. \ref{pic} show that at high enough exciton densities $n_{ex} > 2 \times$ 10$^{12}$ cm$^{-2}$,
there is possible occurrence of a statistically degenerate system  of excitons.
This gives rise to  relaxation mechanisms that are dependent on the  Bose-Einstein statistical
distribution of the excitons.  
In this study, we choose a lower  density $n_{ex}$ = 1$\times$ 10$^{11}$ cm$^{-2}$ so as to exclude degenerate
effects.

Applying Eq.\ref{bec2} to the monolayer MoS$_2$, we obtain
the negative exciton chemical potential $\mu_{ex}$ = -4.8 meV at exciton temperature $T_{ex}$= 20 meV
and the  density,  $n_{ex}$ = 1$\times$ 10$^{11}$ cm$^{-2}$. The chemical potential assumes negative values of magnitude
larger than 5 meV for $T_{ex} > $ 20 meV, hence 
condensation processes can be excluded at typical operating conditions \cite{shi2013exciton,wang2015fast} in  MoS$_2$. 
In Fig. \ref{pic} c, the absolute value of the exciton
chemical potential, $|\mu_{ex}|$ is plotted as a function of the exciton temperature with
exciton  density  fixed at $n_{ex}$ = 1.5 $\times$ 10$^{11}$ cm$^{-2}$ for all the dichalcogenides.
 Molybdenum selenide yields the
highest estimate of $|\mu_{ex}|$ due  to its comparatively large 
exciton mass ($m_e$+ $m_h$). The  selenides show  slightly higher $|\mu_{ex}|$  than the sulfides,
while the Molybdenum  dichalcogenides have higher $|\mu_{ex}|$ than the  Tungsten dichalcogenides.

%%%%%%%%%%%%%%%%%%%%%%%%%%%%%%%%%%%%%%%%%%%%%%%%%%%%%%%%%%%%%%%%%%%%%%%%%%%%%%%%%%%%%%%%%%%%%%%%%%%%%%%

%%%%%%%%%%%%%%%%%%%%%%%%%%%%%%%%%%%%%%%%%%%%%%%%%%%%%%%%%%%%%%%%%%%%%%%%%%%%%%%%%%%%%%%%%%%%%%%%%%%%%%%

\section{Quasi-two dimensional exciton-acoustic phonon
interaction  operator \label{opera}}

The three dimensional electron-phonon
interaction due  to the deformation-potential coupling can be written as 
\be
H^{DF  (3D)}_{e-ph} =
{\sum_{\bf k, q}}
\left(  {\hbar |{\bf q}| \over 2 \varrho \; u V} \right )^{1/2}
\left [  D_c\; a_{1,{\bf k+q}}^{\dagger}\;a_{1,{\bf k}}^{}\;
+ D_v\; a_{0,{\bf k+q}}^{\dagger}\;a_{0,{\bf k}}^{}\;
\right ] \\
\times \left  ( b_{\bf q}^{} + b^{\dagger}_{-\bf q} \right )\;
\label{eop}
\ee
where $D_c$  and $D_v$   are the deformation potentials
for the conduction and valence bands,
$ \varrho$  is the mass density of the bulk system and $u$ is the sound velocity 
of the longitudinal-acoustic phonon mode in the material.
The phonon creation and annihilation operators are indicated by
$b^{\dagger}_{\bf q}$ and  $b_{\bf q}$ respectively.
An explicit form of the  quasi-two dimensional operator can be obtained by
projecting the matrix element of 
the three dimensional electron-phonon interaction operator between two quasi-two dimensional exciton states
  \cite{taka2}. This approach assumes that the electron-hole internal motion
remains unchanged during interaction with the phonons.

Using the   wave  function
in  Eq.\ (\ref{exr}), and the exciton  state vector specified in  
Eqs.\ (\ref{istate}) and (\ref{phi}),  the quasi-two dimensional exciton-phonon interaction
operator can be obtained using  Eq.\ (\ref{eop})
\be
\begin{array}{rcl}
H^{DF  (Q2D)}_{ex-ph} & = & 4 \;
{\sum_{{\bf K_\|, K'_\|},q_z}}
\left(  {\hbar \sqrt { ({(\bf{K_\|-K'_\|})}^2+q_z^2 )}
\over 2 \varrho{_{_m}} \; u{_{_m}} S} \right )^{1/2}
\left [ \frac {D_c\;{\mathcal W}(2 \beta L_w,\; q_z L_w)}{[ 1 + ( {\lambda_h / 2} )^2 ]^{3/2} } -
\frac{D_v\; {\mathcal W}(2 \beta L_w,\; q_z L_w)}{[ 1 + ( {\lambda_e / 2} ) ^2 ]^{3/2} } \right ]  \\
&  &  \qquad\times  | 1s, {\bf K'_\|} \rangle \langle {\bf K_\|},1s |
(b_{{\bf K'_\|-K_\|},q_z}^{} + b^{\dagger}_{{\bf K_\|-K'_\|},q_z} )\;
 \delta_{\bf K-K',q}
\end{array}
\label{exHopr}
\ee
where $\lambda_h={q \alpha_h / \gamma}$ and $\lambda_e ={q \alpha_e/ \gamma}$.
The term $S$ is the surface area of the monolayer plane, $\varrho{_{_m}}$
is the areal mass density and $u{_{_m}}$ is the sound velocity of the phonon mode
in the monolayer system. The  delta  function in Eq.\ (\ref{exHopr}) conserves the momentum of the
scattered exciton and phonon along the $XY$  plane. The function ${\mathcal W}(a,b)$
is given by \cite{taka2}
\be
\label{Gfunc}
G(a,b) = \int_{-1/2}^{1/2} dz_e \int_{-1/2}^{1/2} dz_h \; \exp(i b z_e - a|z_e-z_h|)(1+a|z_e-z_h|)
\cos^2(\pi z_e)\; \cos^2(\pi z_h)
\ee
 The explicit form of the function ${\mathcal W}(a,b)$
is lengthy and we therefore derive ${\mathcal W}(a,b)$ at specific values of 
$a, b$. At the limits $(b \rightarrow 0)$ and $(a \rightarrow 0)$ the  function ${\mathcal W}(a,b)$ take 
simple forms
\bea
{\mathcal W}(a,b \rightarrow 0) &=& \left[3 e^a a^7+28 \pi ^2 e^a a^5+16 \pi ^4 \left(a+(6 a-7) e^a+7\right) a^2+64 \pi ^6 \left(a+(2 a-3) e^a+3\right)\right] 
\nonumber \\
& \times &   \frac{e^{-a}}{2 a^2 \left(a^2+4 \pi ^2\right)^3}
\label{I1}
\\
{\mathcal W}(a\rightarrow 0,b) &=& \frac{2 \pi ^2 \sin \left(\frac{b}{2}\right)}{b(4 \pi ^2 -b^2)}
\label{I2}
\eea

The ideal two  dimensional electron-phonon
interaction for the deformation-potential coupling can be obtained using Eqs. \ref{exHopr}, \ref{I1} and
\ref{I2} as
\be
\begin{array}{rcl}
H^{DF  (2D)}_{ex-ph} & = &  {\sum_{\bf K_\|, K'_\|}}
\left(  {\hbar \sqrt { ({(\bf{K_\|-K'_\|})}^2)}
\over 2 \varrho{_{_m}} \; u_{{_m}} S} \right )^{1/2}
\left [  \frac{D_c}{[ 1 + ( {\lambda_h / 2} )^2 ]^{3/2} }-
\frac{D_v}{ [ 1 + ( {\lambda_e / 2} ) ^2 ]^{3/2} } \right ]  \\
&  &  \qquad\times  | 1s, {\bf K'_\|} \rangle \langle {\bf K_\|},1s |
(b_{\bf q}^{} + b^{\dagger}_{-\bf q} )\;
 \delta_{\bf K-K',q}
\end{array}
\label{exact}
\ee

\section{Relaxation of exciton kinetic energy  \label{relaxpi}}

At low lattice temperatures ($\le$ 30 K), the exciton relaxes by losing its kinetic energy
along the dispersion energy curves coupled with the emission of  acoustic phonons.
The overall rate at which  phonons are emitted is dependent on the energy
exchanges between  the exciton and phonon  during the intra-valley relaxation process.
The net increase in the number of
acoustic phonons is based on  the emission rate of phonons
 with wave vector  ${\bf q} = ({\bf q_\|},q_z)$ 
\be 
\frac{d N_{\bf q}}{dt} = \frac{4 \pi M}{\hbar^3} \; \sum_{\bf K_\|}  \int_{0}^{\pi}  \frac{|H^{DF  (Q2D)}_{ex-ph}({\bf q_\|},q_z)|^2}
{|{\bf K_\|.q_\|} \sin(\theta)|} \delta(\theta-\theta_0) \; \left [(1+{\overline n_{q}}) \;
f({\bf K_\|}+{\bf q_\|})-{\overline n_{q}} f({\bf K_\|}) \right] d\theta,
\label{phoeq}
\ee
where the  thermalized average occupation of phonons 
${\overline n_{q}} = [\exp({\hbar \omega(q) \over k_{_B} T})-1]^{-1}$, 
and $\hbar \omega(q)$ is the energy of phonon with
wavevector $q$. The term $f({\bf K_\|})$ denotes the distribution function associated with 
the exciton wave vector ${\bf K_\|}$. During exciton scattering, 
the energy conservation rule $|{\bf K_\|}| \ge K_l$ is obeyed, where $K_l = |\frac{2 M \omega}{\hbar} - {\bf q_\|}^2|$.
The angle between  the exciton wave vector
${\bf K_\|}$ and ${\bf q_\|}$ is denoted by $\theta_0$.

The relation in Eq. \ref{phoeq} can be further simplified by assuming the action of comparatively large phonon
wave vectors such that ${\bf q_\|} \gg q_m$ where $q_m = {2 M u{_{_m}}}/{\hbar}$ to the following form
\be 
\frac{d N_{\bf q}}{dt} = \frac{ M^{3/2} L^2}{ \sqrt{2} \pi \hbar^4} \; 
  \frac{|H^{DF  (Q2D)}_{ex-ph}({\bf q_\|},q_z)|^2}
{|\bf q_\||} \; \int_{E_l}^{\infty} \; \frac{dE}{\sqrt{(E-E_l)}} \; \left [(1+{\overline n_{ q}}) \;
f(E+\hbar \omega(q))-{\overline n_{q}} f(E) \right],
\label{phoeq}
\ee
where $E_l = \frac{\hbar^2 q_\|^2}{8 M}$. 
Substituting Eqs.\ref{boltz} and \ref{tnum} into Eq.\ref{phoeq}, we obtain an expression for the
average rate of exciton energy relaxation
\bea
\label{relax}
\mean{\frac{dE}{dt}} &=& -{\frac{1}{N}}\; \sum_{{\bf q_\|},q_z} \hbar \omega(q)\; \frac{d N_{\bf q}}{dt} \\
&=& \frac{ \sqrt{2 \pi  M }}{\sqrt{  k_b T_{ex}} \hbar^2  } \; \sum_{{\bf q_\|},q_z} \hbar \omega(q)\;
  \frac{|H^{DF  (Q2D)}_{ex-ph}({\bf q_\|},q_z)|^2}
{|\bf q_\||}\;  \exp \left (-\frac{E_l}{K_b T_{ex}} \right ) \left [(1+{\overline n_{ q}}) \;
 \exp \left(-\frac{\hbar \omega(q)}{K_b T_{ex}} \right)-{\overline n_{q}} \right]
\eea

To  obtain an explicit expression for the average rate of exciton energy relaxation in 
  Eq.\ref{relax}, we employ
the form of  the  quasi- two dimensional exciton-phonon interaction term in Eq.\ref{exHopr}
\bea
\label{relax2}
\mean{\frac{dE}{dt}} &=& - \frac{ \sqrt{2 \pi  M }}{\pi^2 \sqrt{  k_b T_{ex}} \varrho{_{_m}}}
\; \sum_{q_z} \; \int_{q_m}^{\infty} \;dq \; q^2 \;  \exp \left (-\frac{E_l}{K_b T_{ex}} \right ) \left [(1+{\overline n_{ q}}) \;
 \exp \left(-\frac{\hbar \omega(q)}{K_b T_{ex}} \right)-{\overline n_{q}} \right]
\\  \label{relax3} \;\;
& \times &  \left [ \frac {D_c\;{\mathcal W}(2 \beta L_w,\; q_z L_w)}{[ 1 + ( {\lambda_h / 2} )^2 ]^{3/2} } -
\frac{D_v\; {\mathcal W}(2 \beta L_w,\; q_z L_w)}{[ 1 + ( {\lambda_e / 2} ) ^2 ]^{3/2} } \right ]^2
\eea
where $q_m = {2 M u{_{_m}}}/{\hbar}$.
To simplify the numerical evaluation of $\mean{\frac{dE}{dt}}$, we use
$q_z = \frac{\pi}{L_w}$,  which  does not affect
the order of magnitude of the energy relaxation rate. 
 Using the explicit form for   ${\mathcal W}(a,2 \pi)$  derived using  Eq.\ref{Gfunc}, we numerically
evaluate $\mean{\frac{dE}{dt}}$ for the four monolayer 
  transition metal dichalcogenides, MoS$_2$,  MoSe$_2$, WS$_2$ and WSe$_2$
based on  the  material
parameters provided in Table-I. 

\begin{figure}[htp]
  \begin{center}
    \subfigure{\label{figa}\includegraphics[width=6.55cm]{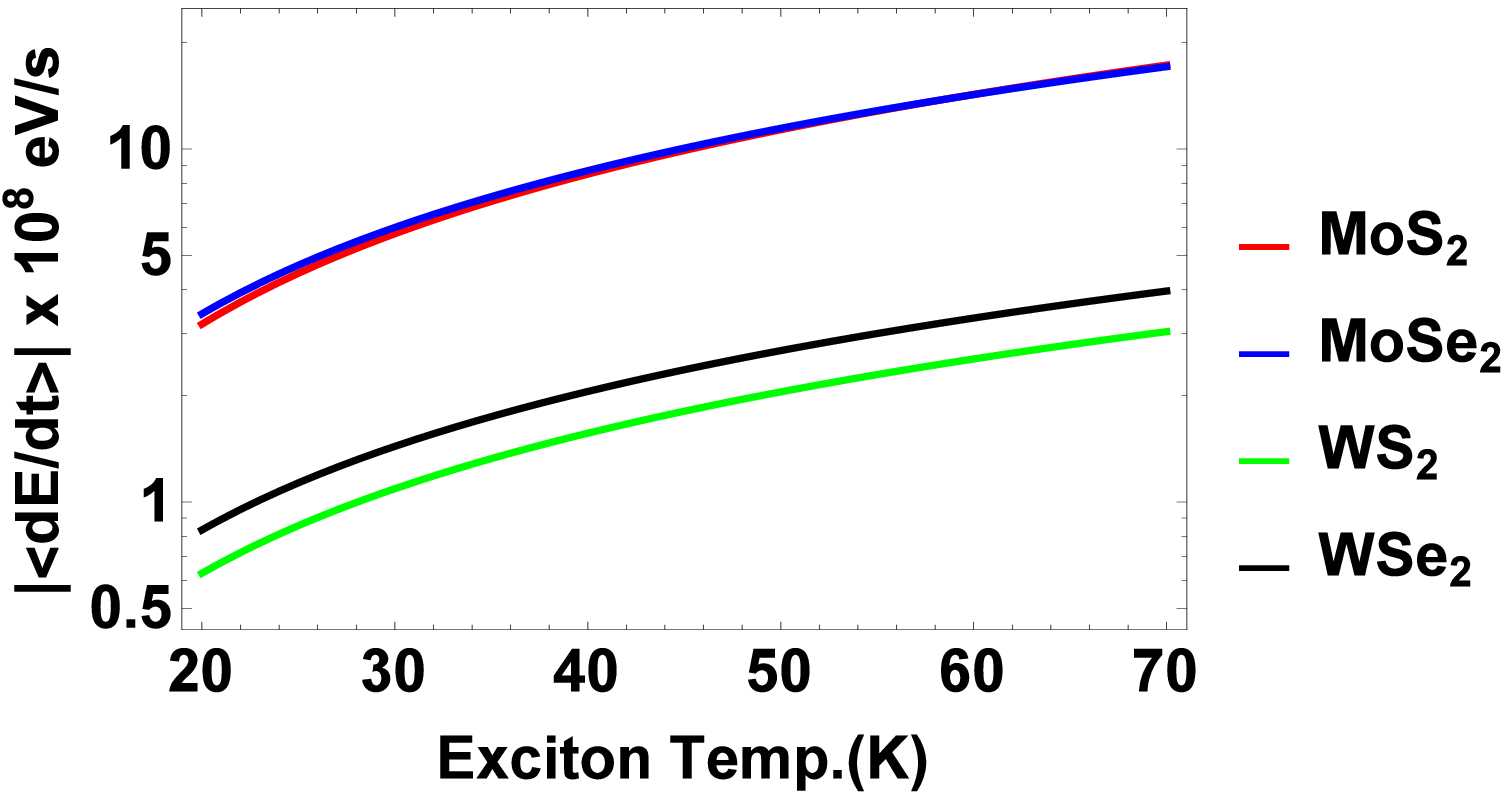}}\vspace{-1.1mm} \hspace{1.1mm} 
\subfigure{\label{figb}\includegraphics[width=4.95cm]{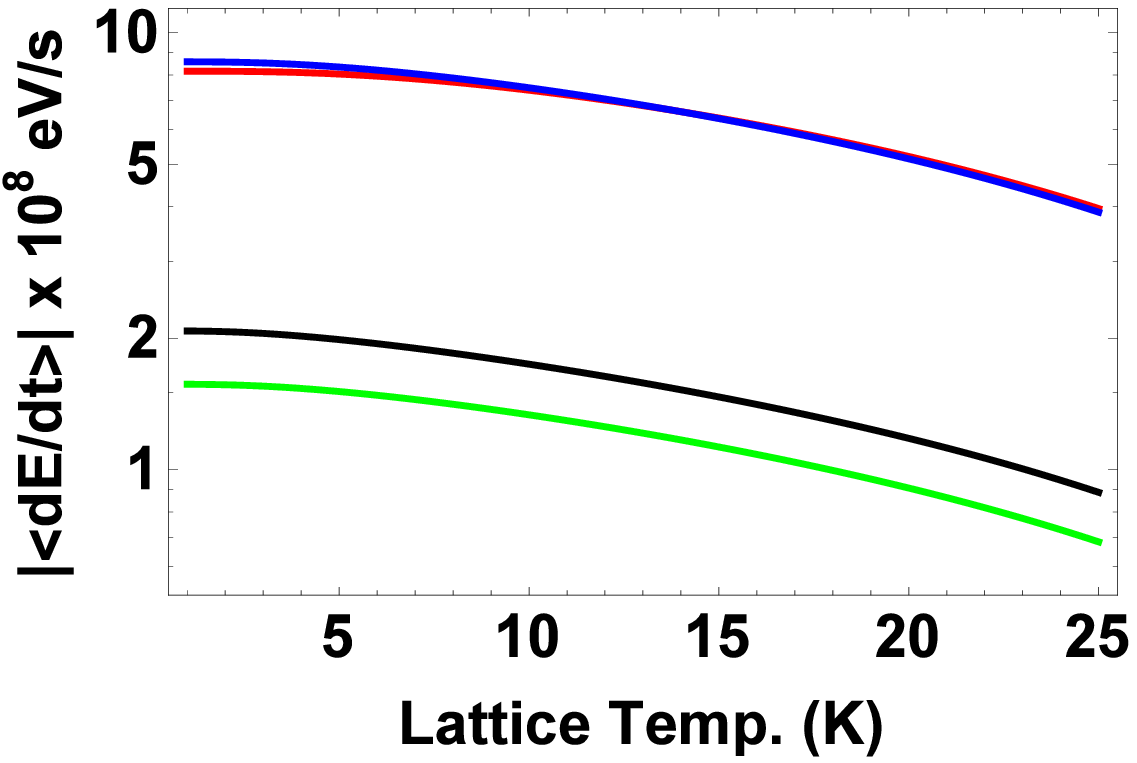}}\vspace{1.1mm} \hspace{1.1mm} 
\subfigure{\label{figb}\includegraphics[width=4.95cm]{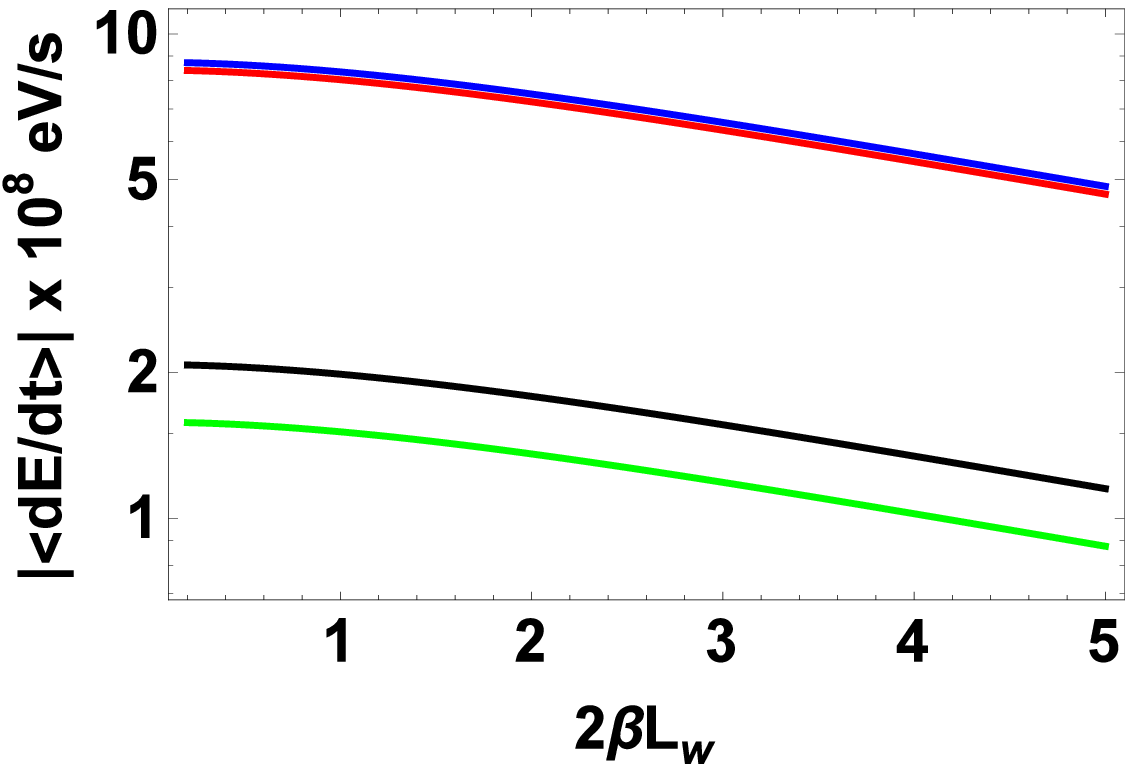}}\vspace{-1.1mm} \hspace{1.1mm} 
 \end{center}
     \caption{(a) Mean relaxation rate  $\mean{\frac{dE}{dt}}$  as function of the 
 exciton temperature $T_{ex}$ in which
 the lattice temperature,  $T_L$ = 5 K and $\beta L_w$ = 0.25
for  monolayer   transition metal dichalcogenides, MoS$_2$,  MoSe$_2$, WS$_2$ and WSe$_2$.
(left), (b) Mean relaxation rate  $\mean{\frac{dE}{dt}}$  as function of the 
 lattice temperature with  exciton temperature $T_{ex}$ = 40K and $\beta L_w$ = 0.25, (c)
 Mean relaxation rate  $\mean{\frac{dE}{dt}}$  as function of parameter, $2 \beta L_w$
evaluated using exciton temperature $T_{ex}$ = 35K
and lattice temperature = 5 K (right)  }
 \label{ratex}
\end{figure}

Fig. \ref{ratex}a shows the increase in  $\mean{\frac{dE}{dt}}$  with exciton temperature $T_{ex}$
for various monolayer   transition metal dichalcogenides  
at  a given lattice temperature $T_L$ = 5 K and confinement parameter $ \beta L_w$ = 0.25.  
The energy relaxation rates for MoS$_2$ lies in the range
$10^{8} -10^{9} $ eV/s for exciton temperatures less than 60 K.
The phonon induced relaxation rates when extrapolated to higher temperatures
with assumption of the Maxwellian distribution for excitons, 
match the experimental estimates
of about $10^{12}$ eV/s
observed at room temperatures.
The experimental results were previously attributed  to defects assisted
intra-band scattering processes \cite{shi2013exciton}. 
We point out that only the relaxation channel via LO-phonons  has been included to evaluate the results
in Fig. \ref{ratex}a. We expect the additional channels provided by TO-phonons,
as well scattering via  piezoelectric coupling (Section \ref{pie}) 
 to further enhance exciton relaxation rates in monolayer systems.
The results in Fig. \ref{ratex} highlight the critical role
played by acoustic phonons  in inducing exciton relaxation processes and 
imply that exciton-phonon interactions can become as strong as exciton-defect interactions
under favourable conditions.

 Fig. \ref{ratex}a shows that the Molybdenum dichalcogenides   experience
 the fastest energy relaxation due to their
high exciton  effective masses,  low mass
densities and high deformation potential constants.
Due to similar material properties, the estimated $\mean{\frac{dE}{dt}}$ are approximately
the same  for MoS$_2$ and MoSe$_2$ monolayer systems. 
Fig. \ref{ratex}b  displays the 
mean relaxation rate  $\mean{\frac{dE}{dt}}$  as function of the 
 lattice temperature at the exciton temperature $T_{ex}$ = 40K and $\beta L_w$ = 0.25.
The results indicate 
a decrease in the effectiveness
of the exciton-acoustic phonon interaction channel as the  lattice temperature
is increased. The parameter $2 \beta L_w$ yields a measure of confinement of
the charge carriers in the direction perpendicular to the monolayer plane.
The decline in  $\mean{\frac{dE}{dt}}$ with increase
in  $2 \beta L_w$ in Fig. \ref{ratex}c shows that reduction in dimensionality
enhances energy relaxation rates in monolayer systems.

%%%%%%%%%%%%%%%%%%%%%%%%%%%%%%%%%%%%%%%%%%%%%%%%%%%%%%%%%%%%%%%%%%%%%%%%%%%%%%%%%%%%%%%%%%%%%%%%%%%%%%%

%%%%%%%%%%%%%%%%%%%%%%%%%%%%%%%%%%%%%%%%%%%%%%%%%%%%%%%%%%%%%%%%%%%%%%%%%%%%%%%%%%%%%%%%%%%%%%%%%%%%%%%

\section {\label{pie} Exciton relaxation due to the  piezoelectric coupling mechanism}

We express the piezoelectric exciton-phonon interaction operator in a form  analogous to
Eq.\ref{exHopr} as   
\be
\begin{array}{rcl}
H^{DF  (Q2D)}_{ex-ph} & = & \frac{4 \;  e }{\epsilon_0}
 \;
{\sum_{{\bf K_\|, K'_\|},q_z}}
\left(  {\hbar \sqrt { ({(\bf{K_\|-K'_\|})}^2+q_z^2 )}
\over 2 \varrho{_{_m}} \; u{_{_m}} S} \right )^{1/2}
\left [ \frac {{\mathbf{e}_{_{11}} \;\mathcal W}(\sqrt{4+\lambda_h^2} \beta L_w,\; q_z L_w)}{[ 1 + ( {\lambda_h / 2} )^2 ]^{3/2} } -
\frac{{\mathbf{h}_{_{11}} \;\mathcal W}(\sqrt{4+\lambda_e^2} \beta L_w,\; q_z L_w)}{[ 1 + ( {\lambda_e / 2} ) ^2 ]^{3/2} } \right ]  \\
&  &  \qquad\times  | 1s, {\bf K'_\|} \rangle \langle {\bf K_\|},1s |
(b_{{\bf K'_\|-K_\|},q_z}^{} + b^{\dagger}_{{\bf K_\|-K'_\|},q_z} )\;
 \delta_{\bf K-K',q}
\end{array}
\label{expieH}
\ee
 where $\mathbf{e}_{_{11}}$ ($\mathbf{h}_{_{11}}$)
is the piezoelectric constant for the electron (hole),
 and $\epsilon_0$ is the dielectric constant that is independent of the 
piezoelectric effect. As given in Eq.\ref{exHopr},  $\varrho{_{_m}}$
is the areal mass density,  $u{_{_m}}$ is the sound velocity of the phonon mode
in the monolayer system and $\lambda_h={q \alpha_h / \gamma}$,  $\lambda_e ={q \alpha_e/ \gamma}$.
Due to  the coupling with 
the TA and LA phonons, piezoelectric interactions are  anisotropic in nature.
These anisotropic effects
can be incorporated   by considering the angular mean of the piezoelectric interaction, which
introduces the factor $\frac{1}{2}$ in Eq.\ref{expieH}.

The piezoelectric constant has been  estimated to be $\mathbf{e}_{_{11}} \approx 3 \times 10^{-11}$ C/m for
monolayer MoS$_2$ \cite{kaasbjergacou}, which is an order of magnitude less than the
estimate $3.64 \times 10^{-10}$ C/m provided by Duerloo et. al. \cite{duerlooemergent} (Table-I).
The reasons for the differences in the piezoelectric constant values for MoS$_2$ from the two
known sources \cite{kaasbjergacou,duerlooemergent} remain unresolved.
According to the piezoelectric constants  given in Ref.\cite{duerlooemergent}, the Molybdenum dichalcogenides   have
higher piezoelectric strengths than the Tungsten dichalcogenides as can be seen in Table-I.
So far all known  piezoelectric constants \cite{kaasbjergacou,duerlooemergent} are linked to the 
linear coupling between the electrical polarization induced by the electron and
 strain field within the crystal. There is no
  mention of similar interactions associated with electrical polarization induced by the hole. 
In order to obtain numerical estimates of  the exciton relaxation rates,
we consider the difference $|\mathbf{e}_{_{11}}- \mathbf{h}_{_{11}}|$ = $\eta \times 10^{-10}$ C/m 
where the parameter $\eta$ is varied from 0.1 to 1.5.

%%%%%%%%%%%%%%%%%%%%%%%%%%%%%%%%%%%%%%%%%%%%%%%%%%%%%%%%%%%%%%%%%%%%%%%%%%%%%%%%%%%%%%%%%%%%%%%%%%%%%%%

\begin{figure}[htp]
  \begin{center}
    \subfigure{\label{figa}\includegraphics[width=8.25cm]{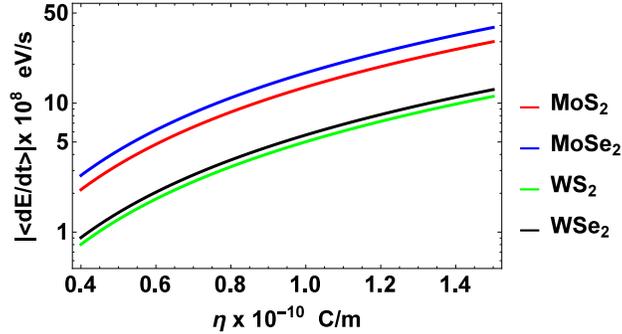}}\vspace{-1.1mm} \hspace{1.1mm} 
 \end{center}
     \caption{ Mean relaxation rate  $\mean{\frac{dE}{dt}}$  as function of the
 difference in piezoelectric constants $|\mathbf{e}_{_{11}}- \mathbf{h}_{_{11}}|$ = $\eta \times 10^{-10}$ C/m. The
exciton temperature $T_{ex}$ = 35 K, and 
 lattice temperature,  $T_L$ = 5 K.}
 \label{ratep}
\end{figure}
%%%%%%%%%%%%%%%%%%%%%%%%%%%%%%%%%%%%%%%%%%%%%%%%%%%%%%%%%%%%%%%%%%%%%%%%%%%%%%%%%%%%%%%%%%%%%%%%%%%%%%

Using Eq.\ref{expieH} and 
 the numerical scheme outlined in Section\ref{relaxpi},
we evaluate the   average rate of exciton energy relaxation $\mean{\frac{dE}{dt}}$
as a function of the difference in piezoelectric constants, $\eta \times 10^{-10}$ C/m.
The results evaluated by setting
exciton temperature $T_{ex}$ = 35 K, and 
 lattice temperature,  $T_L$ = 5 K are displayed  in  Fig. \ref{ratep} 
for four  monolayer   transition metal dichalcogenides.
In all materials, there is increase of $\mean{\frac{dE}{dt}}$ with $\eta$,
and values for $\mean{\frac{dE}{dt}}$ lie in the range
$10^{8} -10^{9} $ eV/s for $|\mathbf{e}_{_{11}}- \mathbf{h}_{_{11}}|$ $\approx 10^{-10}$ C/m.
The order of  rates evaluated in  Fig. \ref{ratep}  are comparable to results 
in   Fig. \ref{ratex}  computed previously for
exciton-phonon interaction due to the deformation potential mechanism. These results indicate
that a unified
approach for exciton-phonon interactions taking into account  both deformation potential
and piezoelectric mechanisms will  yield a higher rate of exciton relaxation than those shown in Figs \ref{ratex}, \ref{ratep}.
The results obtained in this study imply that exciton relaxation induced by  phonons can become as important 
as defects assisted
intra-band scattering processes \cite{shi2013exciton}. 
The role of exciton-exciton interactions during energy relaxation has
not been considered in this work. An earlier work \cite{thilexci}
has shown that   inter-excitonic interactions are sensitive
to changes in the excitonic wavefunctions, indicating that such
interactions may possibly be enhanced
 in monolayers compared to   bulk systems. This can be attributed
 to the enhanced Coulomb coupling between electrons and holes.
A detailed study of the role of inter-excitonic interactions on 
exciton-phonon interactions is beyond the scope of this work as
this requires a refinement in the form of the variational envelope function
employed in Eq.\ref{exr}.
Nevertheless it would be worthwhile to examine whether exciton-exciton interactions 
will influence the phonon assisted exciton relaxation in future studies.

\section{Conclusion \label{con}}
The confinement of excitons  to a narrow region of
space is an important property of monolayer structures
that forms the basis for high excitonic binding energies
and other desirable properties.
In this work  we have examined the relaxation of  quasi-two dimensional 
excitons  due  to interactions with acoustic phonons  via the deformation potential
mechanism.  The   influence of piezoelectric
coupling  linked to electrostatic  interaction  between  the  acoustic
 phonons  and the crystal polarisation
field  is also included in this study.
The relaxation rates due to the corresponding
scattering mechanisms are analyzed for 
 common monolayer   transition metal dichalcogenides (MoS$_2$,  MoSe$_2$, WS$_2$ and WSe$_2$).
The results obtained here indicate that exciton relaxation induced by   phonons due to the deformation potential
and piezoelectric coupling mechanisms
are comparable to defects assisted intra-band scattering processes 
 and  trapping of excitons by surface states
 in monolayer   transition metal dichalcogenides. The results also indicate that
 Molybdenum dichalcogenides    undergo
faster exciton energy relaxation  than the Tungsten dichalcogenides.

 The results  obtained in this work have  importance   in the  optimization of
material properties for device applications and for further exploration of new ideas in physics
for development of  innovative optical and sensor devices. 
Future studies of the effects of phonons on the
 tunneling between valence and conduction band in a  p-n junction \cite{tunnel}
 based on MoS$_2$ monolayer
systems are expected to provide useful results relevant for  device operations.
 The  electronic structures of transition metal dichalcogenides are 
complex and highly  sensitive to electric and magnetic fields. To this end, further investigations 
on the maneuverability of the electronic structures and the influence of external fields on exciton-phonon
interaction is expected to provide an improved understanding of the 
origins of desirable  properties  that can be exploited in the 
development of new devices for future industries.

\eject

\begin{table*}
    \caption{\label{tab:gw} 
Parameters used to obtain results for Fig.\ref{pic}, Fig.\ref{ratex} and Fig.\ref{ratep}.
The effective electron and hole
masses, $m^\ast_e, m^\ast_h$ (in terms of the free-electron mass)
at the $K$  energy valleys/peak are obtained from
Ref. \cite{jin2014intrinsic}. The lattice constant ($a_s$ (\AA) result 
is read from Ref. \cite{kang2013band} while the layer thickness $h$ is taken from
Ref. \cite{ding2011first}.
The deformation potential constants for electron- acoustic phonon (lowest conduction band)
and hole-acoustic phonon interaction (highest valence band) associated with
 the carrier transition, $K \rightarrow K$ are obtained from
Ref. \cite{jin2014intrinsic}. The sound velocities  $u_{_{\rm LA}}$ of the longitudinal acoustic phonon mode
 are derived 
from Ref. \cite{jin2014intrinsic}.  
We evaluate  the monolayer  areal mass density using $\varrho{_{_m}}$ = $\rho \times h$
where the  bulk density $\rho$ is retrieved using the ChemicalData database linked to the Mathematica software
package. For the examined dichalcogenides, the bulk $\rho$ =  5 $\times 10^3$ kg/m$^3$ (MoS$_2$), 
6 $\times 10^3$ kg/m$^3$  (MoSe$_2$),
7.5 $\times 10^3$ kg/m$^3$ (WSe$_2$) and 9.2 $\times 10^3$ kg/m$^3$  (WS$_2$). The estimated
$\varrho{_{_m}}$ indicated in Table-1 show the same order of magnitude $\rho \sim 10^{-7}$ g cm$^{-2}$
for all materials.  
The piezoelectric tensor coefficient $\mathbf{e}_{_{11}}$  for the various monolayer dichalcogenide
 is taken
from Ref.\cite{duerlooemergent}.
In the last two columns of Table-I, we evaluate the dimensionless quantity $q_{_{o}} = \frac{2 M u{_{_m}}}{\hbar} \times 
\frac{a_s}{2 \pi} $ and the exciton bohr radius using 
$a_B$ = 0.529 $\frac{\epsilon}{\mu}$ \AA \; where $\mu$ 
is the exciton reduced mass with $\mu^{-1} = m_e^{-1}+m_h^{-1}$. 
The effective dielectric constant $\epsilon$ for each material is 
based on the estimates used in Ref.\cite{thiljap}. \newline }
\begin{tabular}{|c|c|c|c|c|c|c|c|c|c|c|c}
 %   \begin{tabular*}{\textwidth}{@{\extracolsep{\fill}}ccccccccc@{}}
        \hline
        \hline
System		 ~ & $m_e$, \; $m_h$ & $a_s$ (\AA)  & $h$ (\AA)& $D_c^{op}$ (eV) & $D_v^{op}$(eV)  &
$u_{_{\rm LA}}$$ \times 10^5$ cm/s& $\varrho{_{_m}}$$\times $ $10^{-7}$g/cm$^{2}$ &$\mathbf{e}_{_{11}}$ $\times $ $10^{-10}$C/m & $q_{_{o}}$
 & $a_B$ \AA
\\
        \hline
MoS$_2$ & 0.51, \; 0.58  &  3.18 &  3.13 & 4.5 & 2.5 &6.6 & 1.56 &3.64& 0.006 &7.5  \\
 MoSe$_2$ & 0.64, \; 0.71 &  3.32 &  3.35 & 3.4 & 2.8 &4.1 & 2.01 & 3.92 & 0.005 & 8.0\\
      WS$_2$ & 0.31, \; 0.42 & 3.18 & 3.14   & 3.2  & 1.7 &4.3& 2.36 & 2.71 & 0.003  & 10.6\\
 WSe$_2$ & 0.39, \; 0.51 &  3.32 & 3.36 &  3.2  & 2.1 &3.3& 3.09 & 2.47 & 0.003 &8.6  \\
        \hline
        \hline
    \end{tabular}
    %\end{ruledtabular}
\end{table*}

\end{document}